\documentclass[prd,showpacs,nofootinbib]{revtex4}
\usepackage{amsmath}
\usepackage{amsfonts}
\usepackage{amssymb}
\usepackage{bm}

\begin{document}

\title{Finite Temperature behavior of the CPT-even and parity-even
electrodynamics of the Standard Model Extension}
\author{Rodolfo Casana, Manoel M. Ferreira Jr, Josberg S. Rodrigues and
Madson R. O. Silva}
\affiliation{Departamento de F\'{\i}sica, Universidade Federal do Maranh\~{a}o (UFMA),
Campus Universit\'{a}rio do Bacanga, S\~{a}o Lu\'{\i}s - MA, 65085-580,
Brasil}

\begin{abstract}
In this work, we examine the finite temperature properties of the
CPT-even and Lorentz-invariance-violating (LIV) electrodynamics of
the standard model extension, represented by the term $W_{\alpha \nu
\rho \varphi }F^{\alpha \nu }F^{\rho \varphi }$. We begin analyzing
the hamiltonian structure following the Dirac's procedure for
constrained systems and construct a well-defined and gauge invariant
partition function in the functional integral formalism. Next, we
specialize for the non-birefringent coefficients of the tensor
$W_{\alpha \nu \rho \varphi }$. In the sequel, the partition
function is explicitly carried out for the parity-even sector of the
tensor $W_{\alpha \nu \rho \varphi }$. The modified partition
function is a power of the Maxwell's partition function. It is
observed that the LIV coefficients induce an anisotropy in the black
body angular energy density distribution. The Planck's radiation
law, however, retains its frequency dependence and the
Stefan-Boltzmann law keeps the usual form, except for a change in
the Stefan-Boltzmann constant by a factor containing the LIV
contributions.
\end{abstract}

\maketitle

\section{Introduction}

The researches about Lorentz and CPT violation are commonly performed under
the framework of the standard model extension (SME) developed by Colladay
and Kostelecky \cite{Kostelecky}. The SME is an enlarged version of the
usual standard model that embraces all Lorentz-invariance-violating
coefficients whose tensor contractions yield Lorentz scalars in the observer
frame, and in the particle frame are seen as sets of independent numbers. A
strong motivation to study the SME is the necessity to get some information
about underlying physics at Planck scale where both the Lorentz and CPT
symmetries can be broken due to quantum gravity effects such as it is
suggested by string theory \cite{string}. The photon sector of the SME has
been intensively studied in the latest years with a double purpose: the
determination of new electromagnetic effects induced by the LIV interactions
and the imposition of stringent upper bounds for the magnitudes of the LIV
coefficients. Such investigations have connections with the
Carroll-Field-Jackiw electrodynamics \cite{CFJ},\cite{CFJ2}, consistency
aspects \cite{Consistency}, polarization deviations for light traveling over
large cosmological distances \cite{CFJ},\cite{KM1,KM2}, Cerenkov radiation
\cite{Cerenkov}, radiative corrections \cite{Radiative},
electromagnetostatics and classical solutions \cite{Bailey},\cite{Electro},
\cite{PRD2}, \cite{Paulo}, radiation spectrum of the electromagnetic field
and CMB \cite{Casana2, Winder}, photon interactions and quantum
electrodynamics processes \cite{Interactions}, \cite{Inter2},\cite{Inter3},
\cite{Adam2},\ \cite{Klink2}, \cite{Klink3}, and synchrotron radiation \cite%
{Radiation}. For a large and interesting review on the photon sector and
related issues, see Ref. \cite{Kostelec}. Lorentz violation and its
implications have been studied in several diverse respects \cite{General},%
\cite{General2} and also in other theoretical environments \cite{Ted}.

The most general renormalizable form of the Lorentz-covariance-violating
electrodynamics of the SME photon sector can be expressed by the following
Lagrangian density
\begin{equation}
\mathcal{L}=-\frac{1}{4}F_{\alpha \nu }F^{\alpha \nu }-\frac{1}{4}\epsilon
^{\beta \alpha \rho \varphi }\left( k_{AF}\right) _{\beta }A_{\alpha
}F_{\rho \varphi }-\frac{1}{4}W^{\alpha \nu \rho \varphi }F_{\alpha \nu
}F_{\rho \varphi },  \label{cpt-1}
\end{equation}%
where $\epsilon ^{\beta \alpha \rho \varphi }$ is the totally antisymmetric
Levi-Civita tensor ($\epsilon ^{0123}=1$), $F_{\alpha \nu }$ is the
electromagnetic field tensor, $A^{\alpha }$ is the vector potential, $\left(
k_{AF}\right) _{\beta }=\left( 0,\mathbf{k}_{AF}\right) $ has the dimension
of mass and describes a super-renormalizable (dimension 3) coupling, $%
W^{\alpha \nu \rho \varphi }$ is a renormalizable, dimensionless coupling
giving raise to a dimension 4 operator. The tensor $W^{\alpha \nu \rho
\varphi }$ has the same symmetries of the Riemann tensor $\left[ W_{\alpha
\nu \rho \varphi }=-W_{\nu \alpha \rho \varphi },W_{\alpha \nu \rho \varphi
}=-W_{\alpha \nu \varphi \rho },W_{\alpha \nu \rho \varphi }=W_{\rho \varphi
\alpha \nu }\right] $ and a double null trace which yields only 19
independent components.

The term $\epsilon ^{\mu \nu \kappa \lambda }\left( k_{AF}\right) _{\mu
}A_{\nu }F_{\kappa \lambda }$ is CPT-odd and it was first introduced by
Carroll-Field-Jackiw (CJF) \cite{CFJ}, who studied the modifications
produced by this term (including vacuum birefringence) in the classical
Maxwell electrodynamics. It yields a causal, stable, and unitary
electrodynamics only for a purely spacelike background (see work of Adam \&
Klinkhamer in Ref. \cite{Consistency}). The term $W^{\mu \nu \kappa \lambda
}F_{\mu \nu }F_{\kappa \lambda }$, composed of 19 elements, is CPT-even and
was much investigated in Refs. \cite{KM1,KM2,Bailey, PRD2,Paulo,
Inter2,Klink2, Klink3,Kostelec,Kob}. It yields an electrodynamics not
plagued with stability illness.

As the LIV terms alter the light propagation, it is natural to infer that
the thermodynamical properties of the theory is modified as well. In a
recent work \cite{Casana2}, it was investigated the influence of the CFJ
term on the thermodynamics of the Maxwell field, using the usual formalism
of finite temperature field theory \cite{Kapusta}. It was first analyzed the
Hamiltonian structure of the model using the Dirac formalism in order to
define the partition function of this theory without ambiguities. In the
sequel, the LIV corrections induced on the black body Planck distribution
were carried out, properly examined, and related to the cosmic background
radiation (CMB). A similar investigation remains to be done for the CPT-even
photonic sector of the SME, being it the main purpose of the present work.

This paper is organized as follows. In Sec. II, we discuss the Hamiltonian
structure of the CPT-even electrodynamics, investigating the constraints
structure of the theory (by means of the Dirac formalism). In Sec. III, we
write the general partition function of this electrodynamics into the
functional integral formalism by using the constraint structure and the
gauge fixing conditions established in Sec. II. This partition function is
then particularized and explicitly evaluated for the parity-even sector of
the theory. The modified partition function is written as a function of the
usual Maxwell partition function. In Sec. IV, we present our final remarks
discussing the modifications induced on the Maxwell theory and comparing it
with the Carroll-Field-Jackiw model at finite temperature results. In the
Appendix, we evaluate the dispersion relations for the parity-even sector of
the CPT-even electrodynamics, which corroborate the results obtained for the
partition function of this work.

\section{The CPT-even and LIV electrodynamics of the Standard Model
extension \label{sec-2}}

In the present work, we just study the CPT-even and LIV electrodynamics of
the SME, \textbf{so that} we will consider $\left( k_{AF}\right) _{\beta }=0$%
. Therefore, the Lagrangian density given by Eq. (\ref{cpt-1}) is reduced to
\begin{equation}
\mathcal{L}=-\frac{1}{4}F_{\alpha \nu }F^{\alpha \nu }-\frac{1}{4}W^{\alpha
\nu \rho \varphi }F_{\alpha \nu }F_{\rho \varphi },  \label{cpt-2}
\end{equation}%
which yields the following Euler-Lagrange equation for the gauge field%
\begin{equation}
\partial _{\nu }F^{\nu \mu }-W^{\mu \nu \rho \varphi }{}\partial _{\nu
}F_{\rho \varphi }=0.  \label{cpt-3}
\end{equation}

\subsection{The Hamiltonian structure}

In order to accomplish the Hamiltonian analysis of this model, we begin
defining the canonical conjugate momentum of the gauge field as
\begin{equation}
\pi ^{\mu }=-F^{0\mu }-W^{0\mu \rho \varphi }{}F_{\rho \varphi },
\label{cpt-4}
\end{equation}%
with which we can write the fundamental Poisson brackets (PB): $\displaystyle%
\left\{ A_{\mu }\left( x\right) ,\pi ^{\nu }\left( y\right) \right\} =\delta
_{\mu }^{\nu }\delta \left( \mathbf{x-y}\right) $.

From the Eq.(\ref{cpt-4}), it is easy to note that $\pi ^{0}=0$. Such a null
momentum yields a primary constraint $\phi _{1}=\pi ^{0}\approx 0$ (into the
Dirac formalism, the symbol $\approx $ denotes a \textit{weak equality}).
Also, the momenta $\pi ^{k}$ are defined via the following dynamic relation
\begin{equation}
\pi ^{k}=D_{kj}{}F_{0j}-W^{0kjl}{}F_{jl},
\end{equation}%
where the nonsingular and symmetric matrix $D_{kj}$ is defined by
\begin{equation}
D_{kj}=\delta _{kj}-2W_{0k0j}.  \label{Dmatrix}
\end{equation}%
Then, the velocities $\dot{A}_{k}$ are given as%
\begin{equation}
\dot{A}_{k}=\partial _{k}A_{0}+\left( D^{-1}\right) _{kj}\left[ \pi
^{j}+W^{0jmn}{}F_{mn}\right] ,  \label{cpt-6}
\end{equation}%
while the canonical Hamiltonian density is explicitly written as
\begin{equation}
\mathcal{H}_{C}=\frac{1}{2}\left[ \pi ^{k}+W^{0kmn}{}F_{mn}\right] \left(
D^{-1}\right) _{kj}\left[ \pi ^{j}+W^{0jmn}{}F_{mn}\right] +\pi ^{k}\partial
_{k}A_{0}+\frac{1}{4}\left( F_{jk}\right) ^{2}+\frac{1}{4}%
W^{kjlm}F_{kj}F_{lm}.  \label{cpt-7}
\end{equation}%
Following the usual Dirac procedure, we introduce the primary Hamiltonian $%
\left( H_{P}\right) $ by adding to the canonical Hamiltonian all the primary
constraints, $H_{P}=H_{C}+\displaystyle\int \!\!d^{3}\mathbf{y~}C\pi ^{0}$,
where $C$ is a bosonic Lagrange multiplier. The consistency condition of the
primary constraint, $\dot{\pi}^{0}=\left\{ \pi ^{0},H_{P}\right\} \approx 0$%
, gives a secondary constraint
\begin{equation}
\phi _{2}=\ \partial _{k}\pi ^{k}\ \approx 0.  \label{cpt-8}
\end{equation}%
It means that the Gauss's law structure is not modified by the CPT-even and
LIV background. Nevertheless, expressing it in terms of the electric and
magnetic fields, we can note the explicit coupling between the electric and
magnetic sectors even in the electrostatic regime \cite{Electro,PRD2,Paulo}.

The consistency condition of the Gauss's law gives $\dot{\phi}_{2}=0$. Thus,
the secondary constraint is automatically conserved and there are no more
constraints in this model. The bosonic multiplier of the primary constraint
remains undetermined, being an evidence for the existence of first-class
constraints. This is verified by computing the PB between the primary and
the secondary constraints:\ $\displaystyle\left\{ \pi ^{0},\partial _{k}\pi
^{k}\right\} =0$. The constraints $\phi _{1}=\pi ^{0}\approx 0$ and $\phi
_{2}=\ \partial _{k}\pi ^{k}\approx 0$ reveal that the CPT-even and LIV
electrodynamics has a similar constraint structure as the Maxwell
electrodynamics.

\subsection{Equations of motion and gauge fixing conditions}

Following the Dirac conjecture, we define the extended Hamiltonian $\left(
H_{E}\right) $ by adding all the first-class constraint to the primary
Hamiltonian,%
\begin{equation}
H_{E}=H_{C}+\int d\mathbf{y~}\left[ C\phi _{1}+\Lambda \phi _{2}\right] .
\label{cpt-10}
\end{equation}%
Under this Hamiltonian, we compute the time evolution of the field variables
of the system
\begin{eqnarray}
\dot{A}_{0} &=&\left\{ A_{0},H_{E}\right\} =C,  \label{cpt-11} \\[0.3cm]
\dot{A}_{k} &=&\left\{ A_{k},H_{E}\right\} =\left( D^{-1}\right) _{kj}\left[
\pi ^{j}+W^{0jmn}{}F_{mn}\right] +\partial _{k}A_{0}-\partial _{k}\Lambda ,
\label{cpt-12}
\end{eqnarray}%
Both equations show that the dynamic of gauge field $A_{\mu }$ remains
arbitrary. However, the second equation is similar to the Lagrangian
equation (\ref{cpt-6}) if and only if $\Lambda =0$. Thus, we should impose a
gauge condition in such a way to fix $\Lambda =0$. As it is well-known, the
Dirac algorithm requires a number of gauge conditions equal to the number of
first-class constraints in the theory. However, those gauge conditions must
be compatible with the Euler-Lagrange equations, such that they should fix $%
\Lambda =0$ and determine the Lagrangian multiplier $C$. The gauge
conditions together with the first-class constraints should form a
second-class set.

From the equation of motion for $A_{0}$,%
\begin{equation}
D_{jk}\partial _{j}\partial _{k}A_{0}-W_{0ijk}{}\partial _{i}F_{jk}-\partial
_{0}\left( D_{jk}\partial _{j}A_{k}\right) =0,  \label{eq-A0a}
\end{equation}%
we set as our two gauge fixing conditions%
\begin{equation}
\psi _{1}=D_{jk}\partial _{j}A_{k}\approx 0,~\ \ \psi _{2}=D_{jk}\partial
_{j}\partial _{k}A_{0}-W_{0ijk}{}\partial _{i}F_{jk}\approx 0.  \label{gfc-1}
\end{equation}

The consistency condition for $\psi _{1}$ gives $D_{jk}\partial _{j}\partial
_{k}\Lambda =0$, which fixes $\Lambda =0$. The consistency condition for $%
\psi _{2}$ gives an equation for the multiplier $C$
\begin{equation}
D_{jk}\partial _{j}\partial _{k}C-W_{0ijk}{}\partial _{i}\dot{F}_{jk}\approx
0\ .
\end{equation}%
Consequently, we have determined all the Lagrange multipliers. Therefore,
the set $\Sigma _{a}=\left\{ \phi _{1},~\phi _{2},~\psi _{1},~\psi
_{2}\right\} $ is a second-class one.

The next step is to compute the Dirac brackets to know the field algebra.
Thus, after a long computation, we find that the non-null Dirac brackets are
\begin{eqnarray}
\left\{ A_{k}\left( x\right) ,\pi ^{j}\left( y\right) \right\} _{D}
&=&\delta _{kj}\delta \left( \mathbf{x-y}\right) +D_{mj}\partial
_{m}^{x}\partial _{k}^{x}\bar{G}\left( \mathbf{x-y}\right) ,  \label{DB-1} \\%
[0.3cm]
\left\{ A_{0}\left( x\right) ,\pi ^{k}\left( y\right) \right\} _{D}
&=&-2W_{0ijk}{}\partial _{i}^{x}\partial _{j}^{x}G\left( \mathbf{x-y}\right)
,  \label{DB-2}
\end{eqnarray}%
where the matrix $D_{jk}$ is given by Eq. (\ref{Dmatrix}) and $\bar{G}\left(
\mathbf{x-y}\right) $ is the Green function for the modified Poisson
equation:
\begin{equation}
D_{jk}\partial _{j}\partial _{k}\bar{G}\left( \mathbf{x-y}\right) =-\delta
\left( \mathbf{x-y}\right) .  \label{cpt-18}
\end{equation}

The Dirac brackets for the physical degree of freedom (\ref{DB-1}) do not
reflect the transverse character of the gauge field, however if we choose ${%
\partial _{k}}A_{k}\approx 0$ as a gauge condition, the DB is reduced to the
usual transverse commutation relation%
\begin{equation}
\left\{ A_{k}\left( x\right) ,\pi ^{j}\left( y\right) \right\} _{D}=\left(
\delta _{kj}-\frac{\partial _{k}^{x}\partial _{j}^{x}}{\nabla _{x}^{2}}%
\right) \delta \left( \mathbf{x-y}\right) .
\end{equation}%
Here, we need to observe that, at quantum level, the transverse character of
the gauge field can be explicitly proven by computing the Ward identity for
its 1PI\ 2-point function, $\Gamma ^{\mu \nu }\left( x-y\right) =\left(
\square g^{\mu \nu }-\partial ^{\mu }\partial ^{\nu }+\xi ^{-1}\partial
^{\mu }\partial ^{\nu }-S^{\mu \nu }\right) \delta \left( x-y\right) $, thus
$\partial _{\mu }\Gamma ^{\mu \nu }\left( x-y\right) =\xi ^{-1}\partial
^{\nu }\delta \left( x-y\right) $.

Under the Dirac brackets, the canonical Hamiltonian (\ref{cpt-7}) reads as
\begin{equation}
H=\int d\mathbf{y}\left\{ \frac{1}{2}E^{k}D_{kj}E^{j}+\frac{1}{2}\mathbf{B}%
^{2}+\frac{1}{4}W^{kjlm}F_{kj}F_{lm}\right\} .  \label{cpt-23}
\end{equation}%
In general, for a sufficiently small $W^{\mu \nu \rho \sigma }$, a
positive-definite Hamiltonian is guaranteed, thus providing a stable quantum
theory and a well-defined partition function associated with the CPT-even
and LIV electrodynamics. Now, we proceed to the computation of the partition
function, performing the analysis of its implications to the black body
radiation problem.

\section{The partition function \label{sec-3}}

The next step is to study the thermodynamical properties of the\ CPT-even
photon sector of the SME. The fundamental object for this analysis is the
partition function. The Hamiltonian analysis performed in the previous
section allows to define the partition function (in a correct way) into the
functional integral representation
\begin{equation}
Z\left( \beta \right) =\int \!\!\mathcal{D}A_{\mu }\mathcal{D}\pi ^{\mu
}\delta \left( \phi _{1}\right) \delta \left( \phi _{2}\right) \delta \left(
\psi _{1}\right) \delta \left( \psi _{2}\right) ~\left\vert \det \left\{
\Sigma _{a}\left( x\right) ,\Sigma _{b}\left( y\right) \right\} \right\vert
^{1/2}\exp \left\{ \int_{\beta }\!\!dx~\left( i\pi ^{\mu }\partial _{\tau
}A_{\mu }-\mathcal{H}_{C}\right) \right\} ,  \label{cpt-24}
\end{equation}%
where $\Sigma _{a}=\left\{ \phi _{1},~\phi _{2},~\psi _{1},~\psi
_{2}\right\} $ is a second-class set formed by the first-class constraints
and the gauge fixing conditions, $M_{ab}\left( x,y\right) =\left\{ \Sigma
_{a}\left( x\right) ,\Sigma _{b}\left( y\right) \right\} $ is the constraint
matrix whose determinant is $\det \left( -D_{jk}\partial _{j}\partial
_{k}\right) ^{4}$. Given the bosonic character of the gauge field, its
functional integration can be performed over all the fields satisfying
periodic boundary conditions in the $\tau -$variable: $A_{\mu }\left( \tau ,%
\mathbf{x}\right) =A_{\mu }\left( \tau +\beta ,\mathbf{x}\right) $. The
short notation $\displaystyle\int_{\beta }dx$ denotes $\displaystyle%
\int_{0}^{\beta }\!\!d\tau \!\!\int \!\!d^{3}\mathbf{x}$, and $\mathcal{H}%
_{C}$ is the canonical Hamiltonian given by Eq. (\ref{cpt-7}), and $\beta
=1/k_{B}T,$\ where $k_{B}$\ is the Boltzmann constant.

By performing the integrations over the canonical conjugate momenta and
doing the following redefinitions: $F_{\tau k}=\partial _{\tau
}A_{k}-\partial _{k}A_{\tau }=-F_{k\tau }$ and
\begin{equation}
W_{0k0j}=-W_{\tau k\tau j}~,\text{ ~\ }W_{0kmn}=iW_{\tau kmn},  \label{W1}
\end{equation}%
we find the partition function for the CPT-even photonic sector of the SME
as
\begin{equation}
Z\left( \beta \right) =N\det \left( -D_{jk}\partial _{j}\partial _{k}\right)
\int \mathcal{D}A_{a}\mathcal{~}\delta \left( D_{jk}\partial
_{j}A_{k}\right) \exp \left\{ \int_{\beta }\!\!dx~-\frac{1}{4}F_{ab}F_{ab}-%
\frac{1}{4}\,W_{abcd}F_{ab}F_{cd}\right\} ,  \label{cpt-28}
\end{equation}%
where $a,b,c,d=\tau ,1,2,3$. This partition function is not explicitly
covariant. However, it is well-known that if the covariance is explicit, the
calculation process becomes more manageable. The procedure to pass from a
non-covariant gauge to a covariant one (like the Lorentz gauge $\partial
_{a}A_{a}=0)$ can be performed using the Faddeev-Popov ansatz. Thus,
choosing the following Lorentz gauge $G\left[ A_{a}\right] =-\xi
^{-1/2}\partial _{a}A_{a}+f$, where $f$ is an arbitrary scalar function and $%
\xi $ is a gauge parameter. Therefore, after some algebra, we find the
partition function to be
\begin{equation}
Z\left( \beta \right) =\int D{A}_{a}~\det \left( \frac{-\square }{\sqrt{\xi }%
}\right) \exp \left\{ \int_{\beta }dx-\frac{1}{2}A_{a}\left[ -\square \delta
_{ab}-\left( \frac{1}{\xi }-1\right) \partial _{a}\partial _{b}+S_{ab}\right]
A_{b}\right\} ,  \label{cpt-32}
\end{equation}%
where $\square =\partial _{a}\partial _{a}=\left( \partial _{\tau }\right)
^{2}+\nabla ^{2}$. We have also defined the symmetric LIV operator%
\begin{equation}
S_{ab}=2W_{acdb}\partial _{c}\partial _{d}=S_{ba}.
\end{equation}

For convenience, we choose the Feynman gauge $\xi =1$. Performing the gauge
field integration, we find
\begin{equation}
Z\left( \beta \right) =\det \left( -\square \right) ~\left[ \det \left(
-\square \delta _{ab}+S_{ab}\right) \right] ^{-1/2}.  \label{cpt-33}
\end{equation}%
It is illustrative to mention that this result is similar to that obtained
for the Carroll-Field-Jackiw electrodynamics, where $S_{ab}=\epsilon
_{acdb}(\kappa _{AF})_{c}\partial _{d}$.

Given the high complexity of the CPT-even term, in order to turn feasible
the explicit evaluation of the partition function, the tensor $W_{acdb}$\
should be specialized for simpler configurations. It is done at zero
temperature in Refs. \cite{KM1,KM2}, from which one knows some useful
parametrization for the tensor $W_{\mu \nu \alpha \beta }$ in terms of four $%
3\times 3$ matrices,\textbf{\ }$\kappa _{DE},\kappa _{HB},$ $\kappa
_{DB},\kappa _{HE}$\textbf{:}

\begin{equation}
\left( \kappa _{DE}\right) ^{jk}=-2W^{0j0k},\left( \kappa _{HB}\right) ^{jk}=%
\frac{1}{2}\epsilon ^{jpq}\epsilon ^{klm}W^{pqlm},\left( \kappa _{DB}\right)
^{jk}=-\left( \kappa _{HE}\right) ^{kj}=\epsilon ^{kpq}W^{0jpq}.  \label{P1}
\end{equation}

The matrices $\kappa _{DE}$ and $\kappa _{HB}$ contain together 11
independent components, while $\kappa _{DB}$ and $\kappa _{HE}$ possess
together 8 components, which sums the 19 independent elements of the tensor $%
W_{acdb}$. Such coefficients can be parameterized in terms of\ four
traceless matrices and one trace element. The parity-odd sector is written as%
\begin{equation}
\left( \widetilde{\kappa }_{o+}\right) _{kj}=\frac{1}{2}(\kappa _{DB}+\kappa
_{HE})_{kj},\text{~}\left( \widetilde{\kappa }_{o-}\right) _{kj}=\frac{1}{2}%
(\kappa _{DB}-\kappa _{HE})_{kj},
\end{equation}%
while the parity-even sector is read in terms of two matrices and one trace
element,
\begin{equation}
\left( \widetilde{\kappa }_{e+}\right) _{kj}=\frac{1}{2}(\kappa _{DE}+\kappa
_{HB})_{kj},~~\left( \widetilde{\kappa }_{e-}\right) _{kj}=\frac{1}{2}%
(\kappa _{DE}-\kappa _{HB})_{kj}-n\delta _{kj},~~n=\frac{1}{3}\text{tr}%
\left( \kappa _{DE}\right) .~~
\end{equation}%
The matrix $\kappa _{o+}$ is antisymmetric while the other three are
symmetric. Ten of the 19 elements of the tensor $W_{\alpha \nu \rho \varphi }
$ (5 belonging to $\widetilde{\kappa }_{o-}$ and 5 to $\widetilde{\kappa }%
_{e+}$) are strongly constrained by birefringence data (at the level of 1
part in 10$^{\text{32}}$) \cite{KM1,KM2,Kob}. From the nine remaining
nonbirefringent coefficients, three are contained in the parity-odd matrix $%
\widetilde{\kappa }_{o+}$. The parity-even sector encloses six elements
(five in the matrix $\widetilde{\kappa }_{e-}$\ and the trace element, $n)$.

The prescriptions (\ref{P1}), taking into account the finite temperature
redefinitions (\ref{W1}), are read as%
\begin{equation}
\left( \kappa _{DE}\right) _{kj}=2W_{\tau k\tau j},~\ \left( \kappa
_{HB}\right) _{kj}=\frac{1}{2}\epsilon _{kpq}\epsilon _{jmn}W_{pqmn},~\
\left( \kappa _{DB}\right) _{kj}=-\left( \kappa _{HE}\right) _{jk}=W_{\tau
kpq}\epsilon _{jpq}.
\end{equation}

We should now carry out the determinant of the operator $\left( -\square
\delta _{ab}+S_{ab}\right) $ for the six non-birefringent components of the
parity-even part of the $W_{acdb}$\ tensor.

\subsection{The parity-even sector}

The parity-even sector is composed of an isotropic component and five
anisotropic components - the elements of matrix $\widetilde{\kappa }_{e-}.$\
We now evaluate the partition function for this sector.

\subsubsection{The isotropic contribution}

We first isolate the isotropic part of the parity-even sector by imposing $%
\left( \widetilde{\kappa }_{e-}\right) _{jk}=0$, retaining only the
component $n.$ The functional\textbf{\ }determinant for the operator $\left(
-\square \delta _{ab}+S_{ab}\right) $ is now given as
\begin{equation}
\det \left( -\square \delta _{ab}+S_{ab}\right) =\det \left( n+1\right) ^{2}
\left[ -\square \right] ^{2}\det \left[ -\square +\frac{2n}{n+1}\nabla ^{2}%
\right] ^{2},  \label{cpt-33a}
\end{equation}%
while the partition function becomes
\begin{equation}
\ln Z\left( \beta \right) =-\text{Tr}\ln \left[ -\square +\frac{2n}{n+1}%
\nabla ^{2}\right] .  \label{cpt-33b}
\end{equation}

We can evaluate the involved trace by writing the gauge field in terms of a
Fourier expansion,
\begin{equation}
A_{a}(\tau ,\mathbf{x})=\left( \frac{\beta }{V}\right) ^{\frac{1}{2}}\sum_{n,%
\mathbf{p}}e^{i(\omega _{n}\tau +\mathbf{x}.\mathbf{p})}\tilde{A}_{a}(n,%
\mathbf{p}),  \label{Fourier1}
\end{equation}%
where $V$ designates the system volume and $\omega _{n}$ are the bosonic
Matsubara's frequencies, $\omega _{n}=\displaystyle\frac{2n\pi }{\beta }$,
for $n=0,1,2,\cdots $.

The contributions of the two modes of the gauge field are expressed as
\begin{equation}
\ln Z\left( \beta \right) =-V\int \frac{d^{3}\mathbf{p}}{(2\pi )^{3}}%
\sum_{m=-\infty }^{+\infty }\ln \beta ^{2}\left[ \left( \omega _{m}\right)
^{2}+\frac{1-n}{1+n}\mathbf{p}^{2}\right] ,  \label{cpt-33c}
\end{equation}%
Here, it should hold $\left\vert n\right\vert <1$\ for yielding a
well-defined partition function. By performing the rescaling $%
p_{i}\rightarrow \displaystyle p_{i}\sqrt{\frac{1+n}{1-n}}$, we obtain
\begin{equation}
\ln Z\left( \beta \right) =\left( \frac{1+n}{1-n}\right) ^{3/2}\ln Z_{A},
\label{Z2}
\end{equation}%
where $Z_{A}$ is the partition function of the Maxwell's electrodynamics,
given by
\begin{equation}
\ln Z_{A}=-\frac{V}{\pi ^{2}}\int_{0}^{\infty }d\omega ~\omega ^{2}\ln
\left( 1-e^{-\beta \omega }\right) =V\frac{\pi ^{2}}{45\beta ^{3}}.
\label{ZA}
\end{equation}%
From (\ref{Z2}), we see that the LIV partition function is obviously\textbf{%
\ }a power of\textbf{\ }$Z_{A},$
\begin{equation}
Z\left( \beta \right) =\left( Z_{A}\right) ^{\alpha \left( n\right) },
\label{Z2a}
\end{equation}%
for\ $\alpha \left( n\right) =\left( (1+n)/(1-n)\right) ^{3/2}.$ With this
result, it is easy to show that both the modified Planck's radiation and the
Stefan-Boltzmann's law of the isotropic sector are those of the Maxwell
electrodynamics multiplied by the factor $\alpha \left( n\right) $. Here,
the energy density per solid-angle element remains isotropic.\textbf{\ }

\subsubsection{The anisotropic contribution}

The anisotropic coefficients of the parity-even sector are represented by
the terms of the matrix $\left( \widetilde{\kappa }_{e-}\right) .$ They can
be isolated by setting $n=0$. For evaluating the functional determinant, we
should express the matrix $\left( \widetilde{\kappa }_{e-}\right) $\ in a
suitable way. As the matrix $\left( \widetilde{\kappa }_{e-}\right) $ is
symmetric and traceless,\ it can be parameterized in terms of two orthogonal
3D vectors,$\ \mathbf{a}$ and $\mathbf{b}$, as%
\begin{equation}
\left( \widetilde{\kappa }_{e-}\right) _{jk}=\frac{1}{2}\left(
a_{j}b_{k}+b_{j}a_{k}\right) ,  \label{Param1}
\end{equation}%
with $\mathbf{a}\cdot \mathbf{b}=0$ \ and$~\det \left( \widetilde{\kappa }%
_{e-}\right) =0.$ Then, the functional determinant of the operator $\left(
-\square \delta _{ab}+S_{ab}\right) $ is
\begin{equation}
\det \left( -\square \delta _{ab}+S_{ab}\right) =\det \left( 1-\frac{1}{4}%
\mathbf{a}^{2}\mathbf{b}^{2}\right) \det \left( -\square \right) ^{2}\det
\left( -\square -\nabla _{+}^{2}\right) \det \left( -\square -\nabla
_{-}^{2}\right) ,
\end{equation}%
where the operators $\nabla _{+}^{2}$\ and $\nabla _{-}^{2}$\ are given as
\begin{eqnarray}
\nabla _{+}^{2} &=&\frac{4\left( \mathbf{a}\cdot \nabla \right) \left(
\mathbf{b}\cdot \nabla \right) +\mathbf{b}^{2}\left( \mathbf{a}\cdot \nabla
\right) ^{2}+\mathbf{a}^{2}\left( \mathbf{b}\cdot \nabla \right) ^{2}}{4-%
\mathbf{a}^{2}\mathbf{b}^{2}},  \label{Z1a} \\
\ \ \nabla _{-}^{2} &=&\left( \mathbf{a}\cdot \nabla \right) \left( \mathbf{b%
}\cdot \nabla \right) .  \label{Z1b}
\end{eqnarray}%
With all these definitions, the partition function becomes%
\begin{equation}
\ln Z\left( \beta \right) =-\frac{1}{2}\ln \det \left[ -\square -\nabla
_{+}^{2}\right] -\frac{1}{2}\ln \det \left[ -\square -\nabla _{-}^{2}\right]
~,  \label{Z4}
\end{equation}%
representing the contributions of the two polarization modes of the gauge
field. Let us observe that if we consider only\textbf{\ }the first order
contribution of the LIV background, we have
\begin{equation}
\nabla _{+}^{2}\approx \left( \mathbf{a}\cdot \nabla \right) \left( \mathbf{b%
}\cdot \nabla \right) ,~\text{\ \ }\nabla _{-}^{2}=\left( \mathbf{a}\cdot
\nabla \right) \left( \mathbf{b}\cdot \nabla \right) .
\end{equation}%
It means that the dispersion relation at first order are the same for both
modes of the gauge field,\ once both modes give the same contribution to the
partition function at first order. For an alternative evaluation of the
dispersion relations, see Appendix. This result is compatible with the
statements of Ref. \cite{Kostelec}.

Again, the functional trace is carried out by means of the Fourier expansion
(\ref{Fourier1}) of the gauge field. The contributions of the two modes of
the gauge field are expressed as
\begin{eqnarray}
\ln Z_{+}\left( \beta \right) &=&-\frac{1}{2}V\int \frac{d^{3}\mathbf{p}}{%
(2\pi )^{3}}\sum_{m=-\infty }^{+\infty }\ln \beta ^{2}\left[ \left( \omega
_{m}\right) ^{2}+\mathbf{p}^{2}+\frac{\mathbf{b}^{2}\left( \mathbf{a}\cdot
\mathbf{p}\right) ^{2}+\mathbf{a}^{2}\left( \mathbf{b}\cdot \mathbf{p}%
\right) ^{2}+4\left( \mathbf{a}\cdot \mathbf{p}\right) \left( \mathbf{b}%
\cdot \mathbf{p}\right) }{[4-\mathbf{a}^{2}\mathbf{b}^{2}]}\right] ,
\label{Z4A} \\
\ln Z_{-}\left( \beta \right) &=&-\frac{1}{2}V\int \frac{d^{3}\mathbf{p}}{%
(2\pi )^{3}}\sum_{m=-\infty }^{+\infty }\ln \beta ^{2}\left[ \left( \omega
_{m}\right) ^{2}+\mathbf{p}^{2}+\left( \mathbf{a}\cdot \mathbf{p}\right)
\left( \mathbf{b}\cdot \mathbf{p}\right) \right] .  \label{Z4B}
\end{eqnarray}

In order to perform the momentum integrations, we consider the
following coordinate system: the vector $\mathbf{a}$ is aligned with
the $x-$\ axis, the vector $\mathbf{b}$\ with the $y-$\ axis, so
that $\mathbf{a\times b}$\ points along the $z-$\ axis. Expressing
the momentum in spherical coordinates $\left[ \mathbf{p}=\omega
\left( \sin \theta \cos \phi ,\sin \theta \sin \phi ,\cos \theta
\right) \right] $, we achieve the following
mode contributions:%
\begin{eqnarray}
\ln Z_{+}\left( \beta \right) &=&-\frac{1}{2}\frac{V}{(2\pi )^{3}}\int
d\Omega \int_{0}^{\infty }d\omega ~\omega ^{2}\sum_{m=-\infty }^{+\infty
}\ln \beta ^{2}\left[ \left( \omega _{m}\right) ^{2}+\omega ^{2}\left( 1+%
\frac{a^{2}b^{2}\sin ^{2}\theta +2ab\sin ^{2}\theta \sin 2\phi }{4-a^{2}b^{2}%
}\right) \right] , \\
\ln Z_{-}\left( \beta \right) &=&-\frac{1}{2}\frac{V}{(2\pi )^{3}}\int
d\Omega \int_{0}^{\infty }d\omega ~\omega ^{2}\sum_{m=-\infty }^{+\infty
}\ln \beta ^{2}\left[ \left( \omega _{m}\right) ^{2}+\omega ^{2}\left( 1+%
\frac{1}{2}ab\sin ^{2}\theta \sin 2\phi \right) \right] ,
\end{eqnarray}%
where $d\Omega =\sin \theta d\theta d\phi $ is the solid-angle element, $%
a=\left\vert \mathbf{a}\right\vert $ and $b=\left\vert \mathbf{b}\right\vert
$. By doing a rescaling of the variable $\omega $\textbf{\ }and performing
the summation, we obtain%
\begin{eqnarray}
\ln Z_{+}\left( \beta \right) &=&\frac{1}{8\pi }\ln Z_{A}\int d\Omega
~\left( 1+\frac{\lambda ^{2}\sin ^{2}\theta +2\lambda \sin ^{2}\theta \sin
2\phi }{4-\lambda ^{2}}\right) ^{-3/2}\ , \\
\ln Z_{-}\left( \beta \right) &=&\frac{1}{8\pi }\ln Z_{A}\int d\Omega \left(
1+\frac{1}{2}\lambda \sin ^{2}\theta \sin 2\phi \right) ^{-3/2}\ ,
\end{eqnarray}%
where we have defined $\lambda =ab.$ Taking into account the outcome of Eq. (%
\ref{ZA}),\textbf{\ }and noting that the angular integrations can be exactly
solved, the partition functions become%
\begin{eqnarray}
\ln Z_{+}\left( \beta \right) &=&\frac{1}{4}\left( 4-\lambda ^{2}\right)
^{1/2}~\ln Z_{A}\left( \beta \right) ,  \label{Z3} \\
\ln Z_{-}\left( \beta \right) &=&\left( 4-\lambda ^{2}\right) ^{-1/2}~\ln
Z_{A}\left( \beta \right) .
\end{eqnarray}%
$\allowbreak $For having a well-defined partition function, the product $%
ab=\lambda $ must be bounded as $0<ab<2$. Remembering that $\ln Z\left(
\beta \right) =\ln Z_{+}\left( \beta \right) +\ln Z_{-}\left( \beta \right)
, $we can also show that the LIV\ partition function can be written as a
power of the Maxwell's one,
\begin{equation*}
Z\left( \beta \right) =\left( Z_{A}\right) ^{\delta \left( \lambda \right) },
\end{equation*}%
with\textbf{\ }%
\begin{equation}
\text{\ }\delta \left( \lambda \right) =\frac{1}{4}\left( 4-\lambda
^{2}\right) ^{1/2}+\left( 4-\lambda ^{2}\right) ^{-1/2}.  \label{Z3a}
\end{equation}

Similarly to the parity-odd case, we observe that the modified Planck's
radiation law and the Stefan-Boltzmann's law are those of the Maxwell
electrodynamics multiplied by the factor $\delta \left( \lambda \right) $.
However, the energy density distribution per solid angle,
\begin{equation}
u\left( \beta ,\Omega \right) \,=\frac{\pi }{120\beta ^{4}}\left[ \left( 1+%
\frac{1}{2}\lambda \sin ^{2}\theta \sin 2\phi \right) ^{-3/2}+\left( 1+\frac{%
\lambda ^{2}\sin ^{2}\theta +2\lambda \sin ^{2}\theta \sin 2\phi }{4-\lambda
^{2}}\right) ^{-3/2}\right] ,\
\end{equation}%
possesses an explicitly dependence on $\phi $ and $\theta $ which reveals a
higher degree of the anisotropy induced by LIV, as it can be shown at
leading order\textbf{\ }
\begin{equation}
u\left( \beta ,\Omega \right) \,=\frac{\pi }{120\beta ^{4}}\left[
\allowbreak 2-\frac{3}{2}\lambda \sin 2\phi \sin ^{2}\theta \right] .
\label{pra-1}
\end{equation}%
The $\lambda $ linear dependence of the energy density may lead to an
attainment of upper-bounds on the $\kappa _{e-}$ parameters using
polarization data of the cosmic microwave background.

\section{Conclusions and remarks}

We have initiated this work establishing the Hamiltonian structure of the
CPT-even sector of the electrodynamics of the SME. The constraint analysis
allows the construction of a well-defined partition function which is given
in (\ref{cpt-33}) for an arbitrary and sufficiently small tensor $W_{abcd}$.
At once, we specialize our analysis for the non-birefringent components of
the parity-even parts, for which we compute exactly the partition function.
The expression (\ref{Z3a}) shows that it is a power of the partition
function of the Maxwell electrodynamics, where the power is a pure function
the LIV parameters. This way, the Planck radiation law retains its usual
functional dependence in the frequency and the Stefan-Boltzmann law remains
the same one, apart from a multiplicative global factor containing the LIV
coefficients. It is observed that the LIV induces an anisotropic angular
distribution for the black body energy density for the anisotropic
parity-even $\left( \widetilde{\kappa }_{e-}\right) $ coefficients. The
anisotropic character of the angular radiation distribution reflects local
energy density variations in relation to the Maxwell pattern induced by
Lorentz violation. Despite such differences, the Stefan-Boltzmann law keeps
the usual temperature behavior. This means that, notwithstanding small local
fluctuations, the global radiation law maintains the $T^{4}-$behavior.

Since the LIV coefficients are constrained by very stringent upper bounds,
the\textbf{\ }lower order non-null LIV contribution for the Maxwell
thermodynamics would give a good information about the thermodynamical
properties of the non-birefringent sector the model. It is observed that the
isotropic contribution gives a linear correction in $n$, whereas the
anisotropic contribution coming from the matrix ${\widetilde{\kappa }}_{e-}$
only is manifest at fourth order, as it is shown by Eq. (\ref{Z3}). Hence,
the pure anisotropic contribution is irrelevant when compared with the
isotropic one.

Moreover, we must highlight the differences between the thermodynamical
properties of the CPT-even and the CPT-odd electrodynamics, first
investigated in Ref. \cite{Casana2}. Such difference stems from the Dirac's
algebra of the physical variables. For the CPT-odd electrodynamics \cite%
{Casana2}, in the Coulomb gauge, we have attained
\begin{eqnarray}
\left\{ A_{k}\left( x\right) ,\pi _{j}\left( y\right) \right\} _{D} &=&-
\left[ \delta _{kj}-\frac{\partial _{k}\partial _{j}}{\nabla ^{2}}\right]
\delta \left( \mathbf{x-y}\right) ,  \label{dblv1} \\
\left\{ \pi ^{k}\left( x\right) ,\pi ^{j}\left( y\right) \right\} _{D} &=&%
\frac{1}{2}\left[ \epsilon ^{0kli}\left( k_{AF}\right) _{l}\frac{\partial
_{i}^{x}\partial _{j}^{x}}{\nabla ^{2}}-\epsilon ^{0jli}\left( k_{AF}\right)
_{l}\frac{\partial _{i}^{x}\partial _{k}^{x}}{\nabla ^{2}}\right] \delta
\left( \mathbf{x-y}\right) .  \label{dblv2}
\end{eqnarray}%
Nevertheless, for the CPT-even case and for the Maxwell electrodynamics, the
Dirac algebra is given only by Eq. (\ref{dblv1}). The noncommutativity of
the physical momenta, expressed in Eq. (\ref{dblv2}), is the fundamental
reason for this sector to have different thermodynamical properties when it
is compared with its CPT-even counterpart. Also, since the background $%
k_{AF} $ is a dimensional parameter, the temperature dependence of the
logarithm of the partition function at order $(k_{AF})^{2n}$ changes as $%
T^{3-2n}$. It has as a consequence that the CPT-odd partition function can
not be expressed as a power of the Maxwell one such as it happens in the
CPT-even case.

\subsection{APPENDIX: Dispersion relations}

In this Appendix, we write the dispersion relations for this CPT-even
electrodynamics as a procedure to confirm the evaluation of the partition
functions. A general evaluation for the dispersion relations may be
developed from Eq. (\ref{cpt-2}) and the matrix prescriptions (\ref{P1}). In
terms of the matrices $\kappa _{DE},\kappa _{DB},\kappa _{HB}$ the
non-homogenous Maxwell equations (in the absence of sources) are
\begin{eqnarray}
\partial _{j}E_{j}+\left( \kappa _{DE}\right) _{ja}{}\partial
_{j}E_{a}-\left( \kappa _{DB}\right) _{ja}\partial _{j}B_{a} &=&0, \\
\partial _{0}E_{k}-\epsilon _{kja}\partial _{j}B_{a}+\left( \kappa
_{DE}\right) _{kj}\partial _{0}E_{j}-\left( \kappa _{DB}\right)
_{kb}\partial _{0}B_{b}-\left( \kappa _{DB}\right) _{ab}\epsilon
_{bkj}{}\partial _{j}E_{a}-\left( \kappa _{HB}\right) _{ab}\epsilon
_{bkj}\partial _{j}B_{a} &=&0~,\text{ }~
\end{eqnarray}%
while the homogenous ones remain the same, $\partial _{0}B_{k}+\epsilon
_{kab}\partial _{a}E_{b}=0,$ $\partial _{a}B_{a}=0.$ The wave equation for
the electric field is%
\begin{eqnarray}
&&\left( \partial _{t}\right) ^{2}E_{k}-\nabla ^{2}E_{k}+\text{tr}\left(
\kappa _{DE}\right) \nabla ^{2}E_{k}+\left( \kappa _{HB}\right)
_{ab}\partial _{a}\partial _{b}E_{k}+\partial _{k}\partial _{a}E_{a}+\left(
\kappa _{DE}\right) _{ka}\left( \partial _{t}\right) ^{2}E_{a}  \notag \\
&&-\text{tr}\left( \kappa _{DE}\right) \partial _{k}\partial
_{a}E_{a}-\left( \kappa _{HB}\right) _{kb}\partial _{b}\partial
_{a}E_{a}+\left( \kappa _{HB}\right) _{ka}\nabla ^{2}E_{a}-\left( \kappa
_{HB}\right) _{ba}\partial _{b}\partial _{k}E_{a}  \notag \\
&&+\left( \kappa _{DB}\right) _{kc}\epsilon _{cba}\partial _{t}\partial
_{b}E_{a}-\left( \kappa _{DB}\right) _{ab}\epsilon _{bkc}{}\partial
_{c}\partial _{t}E_{a}=0.  \label{WEE}
\end{eqnarray}

We now specialize the wave equation (\ref{WEE}) for the parity-even case,
setting $\kappa _{DB}=0.$ We then express the matrices $\kappa _{DE},\kappa
_{HB}$ in terms of the $\widetilde{\kappa }_{e+}$ and $\widetilde{\kappa }%
_{e-}$
\begin{eqnarray}
\left( \kappa _{DE}\right) _{ab} &=&\left( \widetilde{\kappa }_{e+}\right)
_{ab}+\left( \widetilde{\kappa }_{e-}\right) _{ab}+n\delta _{ab}, \\
\left( \kappa _{HB}\right) _{ab} &=&\left( \widetilde{\kappa }_{e+}\right)
_{ab}-\left( \widetilde{\kappa }_{e-}\right) _{ab}-n\delta _{ab}.
\end{eqnarray}%
Birefringence data impose $\left( \widetilde{\kappa }_{e+}\right) _{ab}=0,$
so that
\begin{eqnarray}
&&\left( \partial _{t}\right) ^{2}E_{k}+n\left( \partial _{t}\right)
^{2}E_{k}+\left( \widetilde{\kappa }_{e-}\right) _{ka}\left( \partial
_{t}\right) ^{2}E_{a}-\nabla ^{2}E_{k}+n\nabla ^{2}E_{k}-\left( \widetilde{%
\kappa }_{e-}\right) _{ka}\nabla ^{2}E_{a}  \notag \\
&&-\left( \widetilde{\kappa }_{e-}\right) _{ab}\partial _{a}\partial
_{b}E_{k}+\partial _{k}\partial _{a}E_{a}-n\partial _{k}\partial
_{a}E_{a}+\left( \widetilde{\kappa }_{e-}\right) _{kb}\partial _{b}\partial
_{a}E_{a}+\left( \widetilde{\kappa }_{e-}\right) _{ba}\partial _{b}\partial
_{k}E_{a}=0
\end{eqnarray}%
Retaining only the isotropic component ($n\neq 0,\widetilde{\kappa }_{e-}=0$%
), we have%
\begin{equation}
\left[ \left( 1+n\right) \left( \partial _{t}\right) ^{2}-\left( 1-n\right)
\nabla ^{2}\right] E_{k}+\left( 1-n\right) \partial _{k}\partial _{a}E_{a}=0.
\end{equation}%
Using now the first Maxwell equation, $\left( 1+n\right) \partial
_{a}E_{a}=0,$ we obtain $\left[ \left( 1+n\right) \left( \partial
_{t}\right) ^{2}-\left( 1-n\right) \nabla ^{2}\right] E_{k}=0,$ which in
Fourier space, is reads as
\begin{equation}
\left[ \left( 1+n\right) p_{0}^{2}-\left( 1-n\right) \mathbf{p}^{2}\right]
\tilde{E}_{k}=0.
\end{equation}%
This equation yields the following dispersion relation:%
\begin{equation}
\left( 1+n\right) p_{0}^{2}-\left( 1-n\right) \mathbf{p}^{2}=0.
\end{equation}%
This is the same expression contained in Eqs. (\ref{cpt-33a}, \ref{cpt-33b},%
\ref{cpt-33c}), confirming our previous result:
\begin{equation}
\omega _{\pm }=\pm |\mathbf{p|}\sqrt{\left( 1-n\right) /\left( 1+n\right) }.
\label{omega_even1}
\end{equation}%
Here, we see that the phase velocity associated with the modes of the photon
field is the same, showing explicitly the nonbirefringent character of the
isotropic coefficient of the parity-even sector, which is in full accordance
with the statements of Ref. \cite{Kostelec}. Moreover, we note the existence
of positive and negative frequencies, $\omega _{+}$ and $\omega _{-}.$

We should finally consider the anisotropic components of the parity-even
sector ($n=0,\widetilde{\kappa }_{e-}\neq 0$). In this case, the wave
equation (\ref{WEE}) reads as
\begin{equation}
\left[ \square \delta _{ka}+\left( \widetilde{\kappa }_{e-}\right)
_{ka}\square \right] E_{a}-\left[ \left( \widetilde{\kappa }_{e-}\right)
_{cb}\partial _{c}\partial _{b}\delta _{ka}-\left( \widetilde{\kappa }%
_{e-}\right) _{kb}\partial _{b}\partial _{a}\right] E_{a}=0.
\end{equation}%
In momentum space, we have%
\begin{equation}
\left\{ \left[ p^{2}-\left( \widetilde{\kappa }_{e-}\right) _{cb}p_{c}p_{b}%
\right] \delta _{ka}+\left( \widetilde{\kappa }_{e-}\right) _{kb}\left[
p^{2}\delta _{ab}+p_{b}p_{a}\right] \right\} E_{a}=0.  \label{WEEP}
\end{equation}

We now use the same parametrization of Eq. (\ref{Param1}), where $\mathbf{a}$
and $\mathbf{b}$ are two orthogonal 3D vectors. Then, we have%
\begin{eqnarray}
\left( \widetilde{\kappa }_{e-}\right) _{cb}p_{c}p_{b} &=&\left( \mathbf{%
a\cdot p}\right) \left( \mathbf{b\cdot p}\right) , \\
\left( \widetilde{\kappa }_{e-}\right) _{kc}\left[ p^{2}\delta
_{jc}+p_{c}p_{j}\right]  &=&\frac{1}{2}\left( a_{k}b_{j}+a_{j}b_{k}\right)
p^{2}+\frac{1}{2}a_{k}p_{j}\left( \mathbf{b\cdot p}\right) +\frac{1}{2}%
b_{k}p_{j}\left( \mathbf{a\cdot p}\right) .
\end{eqnarray}%
With it, Eq. (\ref{WEEP}) is read as%
\begin{equation}
M_{kj}E_{j}=0,
\end{equation}%
with%
\begin{equation}
M_{kj}=\left[ p^{2}-\left( \mathbf{a\cdot p}\right) \left( \mathbf{b\cdot p}%
\right) \right] \delta _{kj}+\frac{1}{2}\left( a_{k}b_{j}+a_{j}b_{k}\right)
p^{2}+\frac{1}{2}a_{k}p_{j}\left( \mathbf{b\cdot p}\right) +\frac{1}{2}%
b_{k}p_{j}\left( \mathbf{a\cdot p}\right) .
\end{equation}%
The dispersion relations are obtained from $\det \mathbb{M}=0$. Computing
the determinant, we get%
\begin{equation}
\det \mathbb{M}=\left( 1-\frac{1}{4}\mathbf{a}^{2}\mathbf{b}^{2}\right) p^{2}%
\left[ p^{2}-\left( \mathbf{a\cdot p}\right) \left( \mathbf{b\cdot p}\right) %
\right] \left[ p^{2}-\frac{4\left( \mathbf{a\cdot p}\right) \left( \mathbf{%
b\cdot p}\right) +\left( \mathbf{a\cdot p}\right) ^{2}\mathbf{b}^{2}+\mathbf{%
a}^{2}\left( \mathbf{b\cdot p}\right) ^{2}}{4-\mathbf{a}^{2}\mathbf{b}^{2}}%
\right] .
\end{equation}%
By this way, we attain the exact dispersion relations
\begin{eqnarray}
p^{2} &=&\left( \mathbf{a\cdot p}\right) \left( \mathbf{b\cdot p}\right) , \\
p^{2} &=&[4\left( \mathbf{a\cdot p}\right) \left( \mathbf{b\cdot p}\right)
+\left( \mathbf{a\cdot p}\right) ^{2}\mathbf{b}^{2}+\mathbf{a}^{2}\left(
\mathbf{b\cdot p}\right) ^{2}][4-\mathbf{a}^{2}\mathbf{b}^{2}]^{-1}.
\end{eqnarray}%
These expressions confirm that the partition function for this case is the
one stated in Eqs. (\ref{Z4}, \ref{Z4A},\ref{Z4B}). At leading order, these
dispersion relations are the same one,
\begin{equation}
\omega =\pm \left\vert \mathbf{p}\right\vert \left[ 1+\frac{1}{2}\frac{%
\left( \mathbf{a\cdot p}\right) \left( \mathbf{b\cdot p}\right) }{\left\vert
\mathbf{p}\right\vert ^{2}}\right] ,  \label{omega_even2}
\end{equation}%
\textbf{\ }implying absence of birefringence at leading order such as
stabilished in Ref. \cite{Kostelec}.

\begin{acknowledgments}
R. C. thanks Conselho Nacional de Desenvolvimento Cient\'{\i}fico e Tecnol%
\'{o}gico (CNPq) for partial support, M. M. F. is grateful CNPq and to
FAPEMA (Funda\c{c}\~{a}o de Amparo \`{a} Pesquisa do Estado do Maranh\~{a}o)
for partial support. J. S. R. thanks FAPEMA for full support.
\end{acknowledgments}

\end{document}